\documentclass[iop,apj,twocolumn,english]{emulateapj}
\usepackage[T1]{fontenc}
\usepackage[latin9]{inputenc}
\usepackage{geometry}
\usepackage{textcomp}
\usepackage{graphicx}
\usepackage{esint}
\usepackage{babel}
\usepackage{amsfonts}

\begin{document}

\submitted{Author preprint: accepted for \emph{Astrophysical Journal}, 11-April-2014}
\title{Inbound waves in the solar corona: a direct indicator of Alfvén Surface location}

\author{C.E. DeForest$^{1}$, T.A. Howard$^{1}$, and D.J. McComas$^{2,3}$}

\affil{$^{1}$Southwest Research Institute, 1050 Walnut Street, Boulder,
CO, USA}

\affil{$^{2}$Southwest Research Institute, 6220 Culebra Road, San Antonio,
TX, USA}

\affil{$^{3}$University of Texas at San Antonio, TX}

\begin{abstract}
The tenuous supersonic solar wind that streams from the top of the
corona passes through a natural boundary \textendash{} the Alfvén
surface \textendash{} that marks the causal disconnection of
individual packets of plasma and magnetic flux from the Sun
itself. The Alfvén surface is the locus where the radial motion of the
accelerating solar wind passes the radial Alfvén speed, and therefore
any displacement of material cannot carry information back down into
the corona. It is thus the natural outer boundary of the solar corona,
and the inner boundary of interplanetary space. Using a new and unique
motion analysis to separate inbound and outbound motions in synoptic
visible-light image sequences from the COR2 coronagraph on board the
\emph{STEREO-A} spacecraft, we have identified inbound wave motion in
the outer corona beyond 6 $R_s$ for the first time, and used it to
determine that the Alfvén surface is at least 12 solar radii from
the Sun over the polar coronal holes and 15 solar radii in the
streamer belt, well beyond the distance planned for NASA's upcoming
Solar Probe Plus mission. To our knowledge this is the first
measurement of inbound waves in the outer solar corona, and the first
direct measurement of lower bounds for the Alfvén surface.
\end{abstract}


\section{\label{sec:Introduction}Introduction}

The solar corona is distinguished from the solar wind by dynamical
means. Coronal plasma is, on average, continuously expanding into
interplanetary space to form a fast wind (\citealt{Parker1958,Neugebauer1962})
that forms the heliosphere (\citealt{Parker1961,Axford1963}). In
the corona, the plasma motion is slower than the speed of the MHD
wave modes. In the heliosphere, the plasma is super-Alfvénic, so that
information cannot propagate inward and affect the morphology or connectivity
of the solar corona. The two regions are divided by a boundary, the
\textquotedblleft Alfvén surface\textquotedblright , at which the
wind speed exceeds the Alfvén speed (formally the fast magnetosonic
speed, but we use the term ``Alfvén speed'' throughout this study
as they are equal in the field-aligned direction and the local magnetic
field is nearly radial in the outer corona). This boundary has also
been called by several other names, among them the ``heliobase'' (\citealt{Zhao2010}),
the ``Alfvén point'' (\citealt{Hundhausen1972}), and the ``Alfvén
radius'' (\citealt{Goelzer2014}).

The Alfvén surface is fundamental to the magnetic topology of the
solar corona and heliosphere. Magnetic flux that passes through Alfvén
surface boundary is referred to as \textquotedblleft open\textquotedblright{}
in the context of coronal physics. Because usage of ``open'' has
diverged and become ambiguous across both the solar remote sensing
and heliospheric in-situ sensing communities, we use the phrase \textquotedblleft Alfvén
open\textquotedblright{} to distinguish field lines that cross through
the Alfvén surface, from field lines that pass through other important
surfaces or qualify as ``open'' under other definitions.

In addition to its importance for the corona and for MHD simulations,
the Alfvén surface should be detectable remotely via motions in the
visible corona. Outside the Alfvén surface, all collective motions of
the plasma must propagate outward from the Sun. Inside the Alfvén
surface, such motions may propagate both outward and
inward. \citet{Verdini2009} modeled the speed-vs-radius behavior of
inbound waves, which yield a specific signature of slow inbound
propagation near the Alfvén surface.  It should be possible to
identify the approximate location of the Alfvén surface in different
regions of the corona by examining the spatial spectrum and relative
intensity of inward and outward propagating disturbances in images of
the outer corona, provided that sufficiently low-noise measurements
are available and an analysis technique can be developed to separate
the upward and downward motions. This latter requirement comes from
the fact that, except in special circumstances such
as the retracting of \emph{inner} coronal loops below 6 $R_s$
(\citealt{McKenzie1999,Wang1999,Sheeley2001b,DeForest2012}),
the outward motion is so dominant that it is extremely difficult to
discern any inward motion at all.

Consideration of the Alfvén surface to date has been mostly theoretical (that
is, it has been considered primarily using theory, models and simulations),
with treatments dating back to the division of ideas regarding the
expansion of the corona between \citet{Chapman1957} and \citet{Parker1958}
in the 1950s. \citet{Parker1958} first suggested that there was a
division between the corona and the solar wind, while \citet{Hundhausen1972},
in a review summarizing the theoretical developments of the expanding
corona leading to the early 1970s, described the nature of this division
in terms of the transition from ``closed'' to ``open'' magnetic
field lines. Several recent works suggest a broad range of possible
distances for the Alfvén surface, from 10\textendash 30 $R_{S}$.
\citet{Zhao2010} used observations of helmet streamers in the corona
to model the Alfvén surface, placing its outer limit at 10\textendash 14
solar radii ($R_{S}$) around solar minimum. \citet{Wang1999}, and \citet{Sheeley2001a}, 
and \citet{Sheeley2004} 
have observed isolated inbound retracting loops in streamers; never observing such a 
feature beyond $6 R_{s}$, they attribute this lack to a low
Alfv\'en surface near $6 R_{s}$.  \citet{Schwadron2010}
and \citet{Smith2013} treat the Alfvén surface in the context of
the heliospheric flux balance, and place it at 10-15 $R_{S}$. \citet{Goelzer2014}
have applied a simple model of the heliospheric magnetic field to
in-situ measurements of the solar wind and place the surface around
15 $R_{S}$ at solar minimum and 30 $R_{S}$ at solar maximum. 

In the present work, we report on the first detection and measurement
of inbound wave and other motions in the outer corona, using synoptic
data from \emph{STEREO}/COR2 (\citealt{Howard2008}) and post-processing
to separate image features by characteristic direction of motion.
We have measured signatures of inbound motion, which we attribute
to propagation of compressive waves in the corona (e.g. \citealt{DeForest1998}),
over the full range of altitudes viewed by COR2 in the streamer belt,
and out to 12.5 $R_{S}$ in the polar coronal holes. Based on these
measurements, we conclude that the Alfvén surface is typically above
15$R_{S}$ in the streamer belt and well above 12 $R_{S}$ in the
polar coronal holes in solar minimum conditions. In Section 2
we describe the theory of measurement; in Section 3
we describe the dataset we analyzed and the techniques used to prepare
inbound and outbound images; in Section 4
we present
direct results of the analysis; and in Section 5
we discuss their implications and required follow-on analysis, before
summarizing the work in Section 6.

\section{\label{sec:Theory-of-measurement}Theory of measurement}

To understand the expected visual signature of waves in the outer
corona, we briefly discuss the theory of coronagraphic measurement.
Coronagraphs record Thomson-scattered light from the optically thin
corona. The viewing angles are small and it is customary to approximate
viewing coordinates with a Sun-centered Cartesian coordinate system
(\emph{x,y,s}), where the first two coordinates are scaled from the
image plane and the third is distance along each line of sight. The
local differential radiance of the corona depends on the local electron
density and a geometric function that varies only slowly with \emph{s}
(e.g. \citealt{Billings1966}). Other sources of pixel brightness
include the starfield, F corona, and instrument stray light (e.g.
\citealt{Lyot1939,Brueckner1995}). Through post-processing one typically
eliminates (or at least greatly reduces) these background sources,
so that the processed coronal radiance \emph{B'} is given by:
\begin{equation}
B'(x,y,t)\approx k(r)\int ds\left(n'_{e}(x,y,s,t)\right)\,+\, N'(x,y,t)\label{eq:final}
\end{equation}
where \emph{r} is focal-plane radius from the Sun; \emph{k }is a per-radius
constant of proportionality that includes the instrument calibration
geometric factors, mean solar radiance, and Thomson scattering physics;
$N'$ is a residual noise term, which includes photon statistics and
also unsubtracted background; and $n_{e}'\equiv n_{e}(x,y,s,t)-n_{0}(x,y,s)$
for some baseline time-independent $n_{0}$ that is subtracted as
part of the estimation and removal of the background sources (e.g.
\citealt{Morrill2006}). Because of the steep radial gradient in density
within the corona, the integral in Equation \ref{eq:final} is dominated
by the region where $s\ll r$, i.e. the ``sky plane'' (e.g. \citealt{FisherGuhathakurta1995}).

A ``feature'' or bright patch in $B'$ generally represents a locus
of enhanced density in the solar corona (an ``object'' or ``structure''),
and a moving feature (i.e. one that exhibits displacement in subsequent
images) thus represents either true motion of dense coronal
material, propagation of a compressional wave signal through the coronal
medium, or some combination of these. An additional possible source
of apparent motion is alignment between the line of sight and an extended
structure such as a slightly curved thread or sheet in the
corona; these ``caustic effects'' can in principle cause rapid apparent
motion as a result of slight changes in the position or shape of the
structure. These effects only occur during rare coincidences and we
presume them to be negligible. 

Pure Alfvén waves themselves include no variation of density and hence
are not visible with a coronagraph; but fast-mode MHD waves are
visible and propagate at speeds between $V_{A}$ and
$V_{Fmax}=\sqrt{C_{s}^{2}+V_{A}^{2}}\approx V_{A}$ relative to the
medium, where $C_{s}$ is the speed of sound, $V_{A}$ is the Alfvén
speed, and $C_{s}/V_{A}$ is the plasma $\beta$ parameter, which is
generally small in the corona (e.g. \citealt{Priest1982}).  Above the
lowest layers of the corona, outward wind flow dominates the plasma
motion (e.g. \citealt{Parker1958,Hundhausen1972}), so that measured
inbound motion of features in $B'$ is most frequently caused by wave
action, although retraction of loops, with corresponding plasma
motion, is seen in and around the streamer belts at altitudes below $6
R_s$ (e.g. \citealt{Wang1999,Sheeley2001a,Sheeley2004}).  Note that,
although wave fields are commonly described using the plane wave
basis, wave motion is not required to have any oscillatory character
at all. Wave-related motion or density enhancement can have a smooth,
pulse, complex, or oscillatory character depending on the excitation
and any resonances in the system supporting the waves. In the corona,
we expect to observe waves that are excited by the passage of outbound
coronal structures such as coronal mass ejections, blobs
(\citealt{Sheeley2009}), or disconnected U-loops
(\citealp{McComas1991,DeForest2012}). These waves are needed to carry
the inbound signals that describe and set the new equilbrium shape of
the corona.

Waves propagating in a moving medium are advected with the medium,
so that if the bulk radial wind speed $V_{w,r}$ be significantly
greater than the Alfvén speed $V_{A}$, no inbound features should
be observed at all. But if the bulk radial wind speed $V_{w,r}$ happens
to be significantly less than the Alfvén speed, then inbound features
should be detected - particularly after passage of a CME, blob, or
other localized disturbance that causes a shift in the coronal equilibrium.
Such shifts can only propagate inward at speeds up to $V_{A}-V_{w,r}$,
and should be visible in carefully prepared image sequences, just
as they are in modeled image sequences of the wind acceleration region
(e.g. \citealt{Verdini2009}).

In practice, such features have never (to our knowledge) been observed
beyond a few $R_{S}$ in the coronal holes, though inbound wave signals
must be present if (as is observed) the inner corona reacts to large
scale changes above altitudes of $4-5\, R_{S}$. One reason for this
lack may be that the unaided eye has difficulty separating the presumably-faint
inbound wave signal from a far greater optical flow%
\footnote{Readers are reminded that ``optical flow'' is the pattern of apparent
motion in a visual scene, as distinct from actual flow of structures
imaged in the scene. We use the phrase to refer to the image energy that
is present within a particular range of velocities, as distinct from the 
apparent-to-the-eye motion of individual features in the scene.
} of outbound features. We overcame this difficulty by using Fourier
transformation to separate fully the inbound and outbound features
in the coronagraphic image data. 

Converting $B'$ from Cartesian $(x,y,t)$\emph{ }to polar $(\theta,r,t)$
coordinates yields a movie in which radial motion is purely vertical.
Fourier transformation from $r$ and $t$ to wavenumber $k_{r}$ and
frequency $\omega$ localizes all moving features in the movie, regardless
of size or location, to the line through the origin whose slope is
equal to the radial speed of the feature (see the Appendix).
Figure \ref{fig:Fourier-transformation-localizes} shows the location
of inbound and outbound image features in the $(k_{r},\omega)$ plane
(neglecting the sky-plane azimuthal angle $\theta$ or its inverse,
$k_{\theta}$). By masking out unwanted parts of the $(k_{r},\omega)$
Fourier plane and then analyzing the remaining energy in the images,
we were able to search for inbound features, in the absence of distraction
from the dominant outward motion. 

\begin{figure}
\center{\includegraphics[width=1.75in]{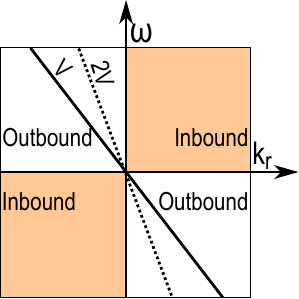}}

\protect\caption{\label{fig:Fourier-transformation-localizes}Fourier transformation
localizes moving features by speed and direction. Features moving
with speed $V$ in the $(r,t)$ plane are transformed to features
lying along the indicated line in the $(k_{r},\omega)$ plane, which
is segmented into inbound and outbound quadrants as marked. Doubling
the speed to $2V$ doubles the slope of the line.}
\end{figure}

Two kinds of non-directional signals are captured by the Fourier quadrant
filter shown in Figure \ref{fig:Fourier-transformation-localizes}.
First, non-moving features such as the streamer belt itself are formed
of equal parts inbound and outbound energy in Fourier space, and therefore
they appear in both inbound and outbound sequences. Fortunately, the
only non-moving features in the corona are also quasi-stationary,
i.e. they exist at low $\omega$ only. These features can be removed
by simply zeroing the low-$\omega$ portion of the dataset, an operation
that is similar to unsharp masking in time. Secondly, isotropic
noise sources such as the photon noise contain an isotropic mix of
wave signals and therefore appear equally in the inbound and outbound
portions of the separated movie. These noise sources are identifiable
precisely because they are nearly isotropic: they can be eliminated
by searching for structure in the filtered data, such as a narrow
range of speeds far from any characteristic speed of the filters that
have been applied. Wave signals in the data are expected to propagate
at the local wave speed corrected for advection, while the only characteristic
speeds in the noise should be any that are imposed by the filtering
and data-preparation process.

\section{\label{sec:Data-=000026-Methods}Data \&\ Methods }

\begin{figure*}
\center{\includegraphics[width=5.0in]{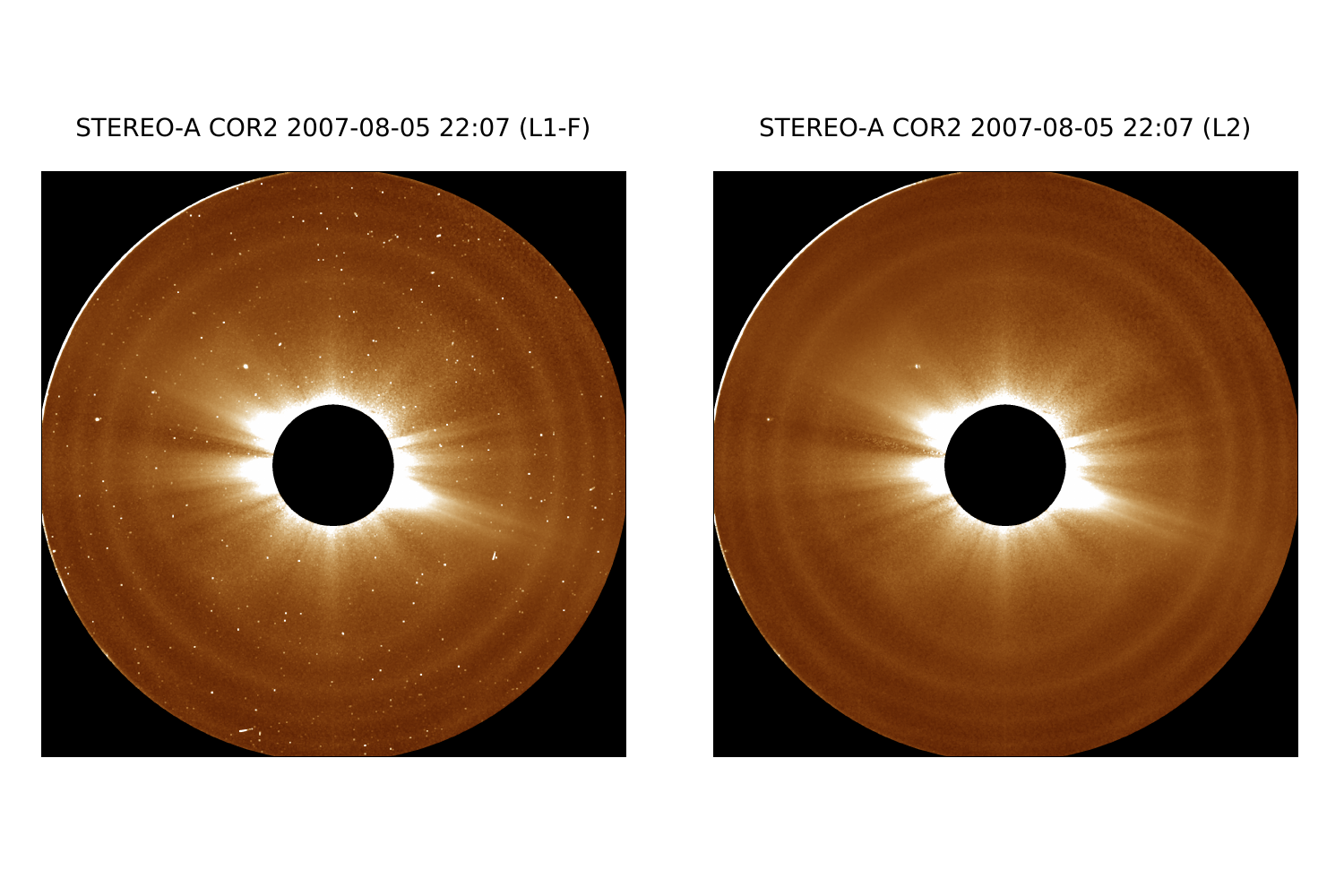}}
\protect\caption{\label{fig:We-prepared-COR2}We prepared COR2 data by generating (and
subtracting) an \emph{ad hoc} F coronal model (left), and then further
despiking the data (right) to remove most stars. }

\end{figure*}

We sought to identify the Alfvén surface in a sequence of coronagraph
images from \emph{STEREO-A}/COR2 (\citealt{RHoward2008}) by searching
for inbound feature motion through a sequence of coronagraph
images. We selected 2007 August 4\textendash 14 as a quiet period near
solar minimum with a small amount of coronal activity and no
instrumental anomalies. We downloaded the Level 0 data from the
\emph{STEREO} web site, processed it to Level 1 with the
\emph{SECCHI\_PREP }program available via Solarsoft
(\citealt{FreelandHandy1998}), and carried out several further
nonstandard steps to improve and regularize the data. First, we
prepared a model F corona by smoothing each Level 1 image by a
9-pixel-diameter tophat kernel,\footnote{Readers are reminded that a
  ``tophat kernel'' is a generalization to two dimensions of the
  familiar ``boxcar kernel'' in one dimension. It is constructed by
  starting with an image containing all zeroes, then setting all
  pixels within a given radius of the center to unity and all pixels
  outside that radius to zero.  Finally, the kernel is normalized to a
  sum of unity.}  
then taking the 1 percentile value (i.e. 5th lowest)
of the 512 values for each pixel in the data set. We subtracted this
model F corona from each frame to produce a K coronal movie; these are
the ``L1-F'' data, and a typical frame is shown in the left panel of
Figure \ref{fig:We-prepared-COR2}.

Because we sought to find wavelike patterns in the data, we needed
to minimize the residual starfield (frequently ignored in coronagraph
analysis). We applied the \emph{spikejones }despiking algorithm (\citealt{DeForest2004a})
to each frame of the L1-F data. This was sufficient to remove most
visible stars but left wide PSF-derived ``halos'' from most bright
stars. To further reduce the effect of the starfield, we considered
each pixel in the dataset as a time series. We generated a smoothed
copy of the time series by applying a 9-frame median filter, and identified
the values in the original data whose difference from this median-smoothed
copy was the greatest. We set the corresponding pixel in each of the
the 10 frames with the highest difference value, to the timeseries
median. This had the effect of removing most data dropouts and most
stellar halo effects from the image sequence. We called these data
Level 2, and a typical frame is shown in the right panel of Figure
\ref{fig:We-prepared-COR2}.

The additional postprocessing with \emph{spikejones} and the median
filter eliminated most stars, but small ``halos'' are still present
around the very brightest objects (such as the planet Mercury and
the brightest few stars). These remain faintly visible in the data
and are highlighted by the subsequent processing steps, but are clearly
identifiable from their slow motion and compact form. 

\begin{figure*}
\center{\includegraphics[width=5.0in]{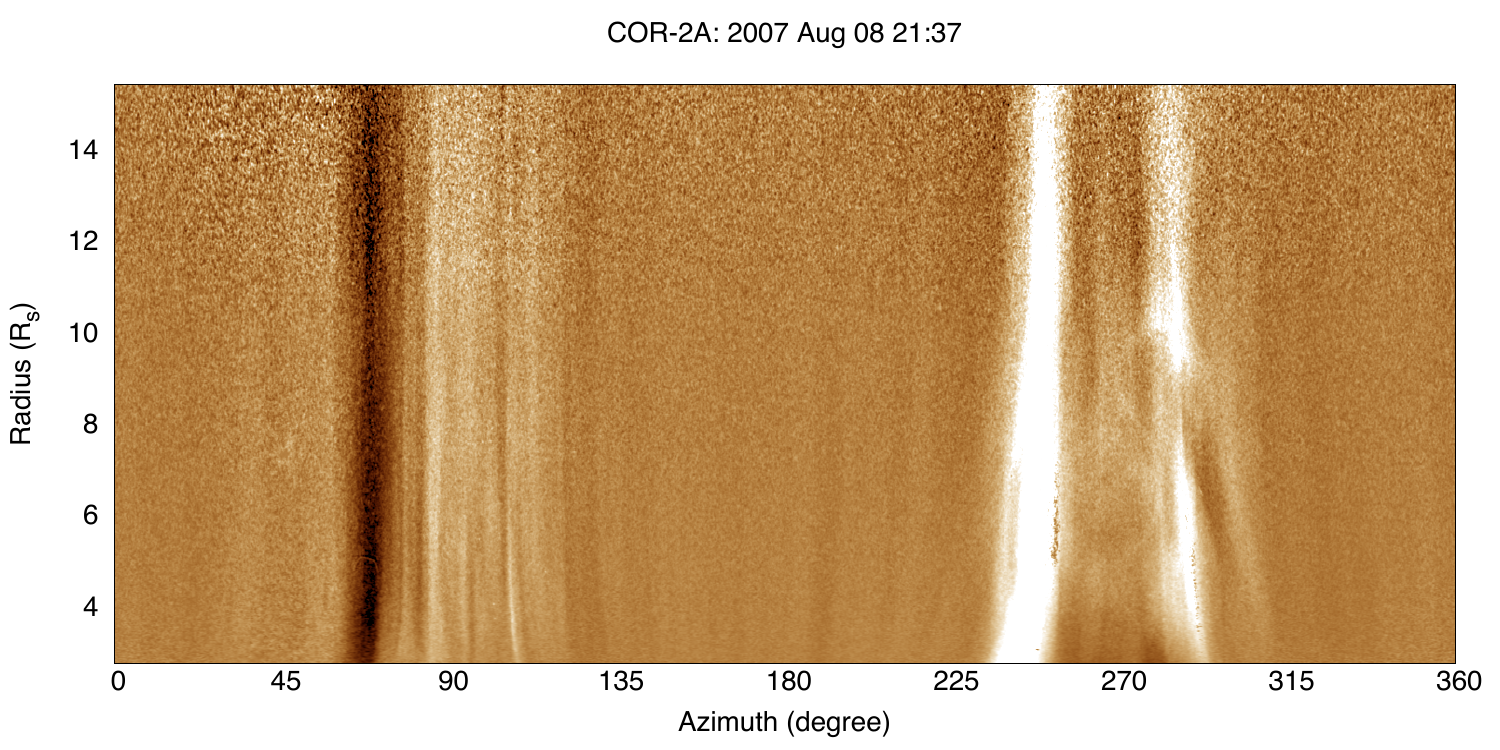}}
\protect\caption{\label{fig:Radialized-COR2-frame}Radialized COR2 frame shows prepared,
radialized images prior to motion filtering.}
\end{figure*}

To give better access to radial motion of differential signals, without
regard to the overall gradient in radiance, we resampled each image
into radial coordinates, and normalized with a radial filter. The
resampling step used a spatially-variable resampling filter to avoid
introducing moiré artifacts (\citealt{DeForest2004b}). We normalized
radially by subtracting from each row its mean value across column
and time, then dividing the row by its variance (RMS value) across
column and time. The result is a radial-coordinate frame such as Figure
\ref{fig:Radialized-COR2-frame}. We transposed the sequence of radial
frames into a collection of (radius, time) evolution images, one at
each of 720 azimuthal angles.

To isolate moving signals, we Fourier transformed the radialized image
sequence in radius and time, and divided the ($k_{r},\omega$) plane
into quadrants to separate the data into inbound and outbound sequences.
At this stage, we also imposed a motion filter, rejecting all features
moving slower than 1 pixel per frame (19 km sec$^{-1}$) and all features
moving faster than 47 pixels per frame (900 km sec$^{-1}$). These
speeds were selected to be broad enough to capture features moving 
between a significant fraction of the sound speed and the Alfv\'en 
speed, but do not themselves hold any particular significance.
The motion filtering
removed the stationary and quasi-stationary streamer signals, making
a motion signal easier to perceive. The theory of motion-filtering 
and our application of it are described in more detail in the Appendix.

\begin{figure*}
\center{\includegraphics[width=5.0in]{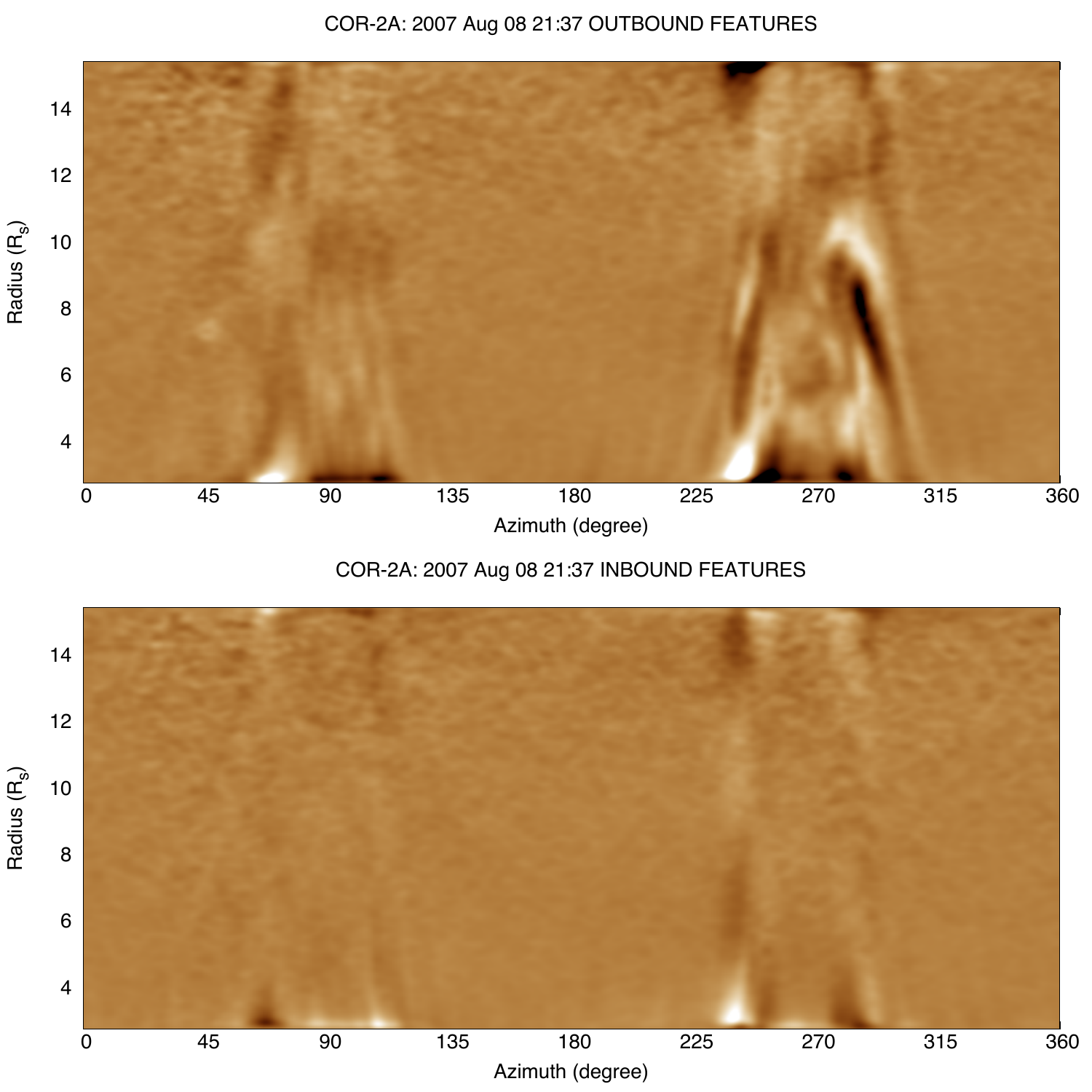}}
\protect\caption{\label{fig:Fully-filtered-COR2}Fully filtered and inbound/outbound
separated COR2 frame shows lower noise and lack of stationary streamer
compared to Figure \ref{fig:Radialized-COR2-frame}. TOP: outbound
features show a CME in progress over the east limb (270\textdegree ).
BOTTOM: inbound features show ringing near the edges of the streamers,
and a weak return signal. The full movie is available as a supplement
in the digital version of this article.}
\end{figure*}

Also we smoothed each radialized
frame by convolution with a 5 pixel full-width Gaussian in the image
plane (i.e. an elliptical Gaussian with 2.5 degrees full-width in
azimuth and 0.2 $R_{s}$ in radius). This further reduced image noise,
especially in the outer portions of the image plane where the original
signal is faint. One of these fully filtered frames is pictured in
Figure \ref{fig:Fully-filtered-COR2}, divided into inbound and outbound
images. The full set of filtered frames is available as a supplementary
movie in the digital version of this article; the respective outward
and inward motions are clearly apparent in the movie.

\begin{figure*}
\center{\includegraphics[width=5.0in]{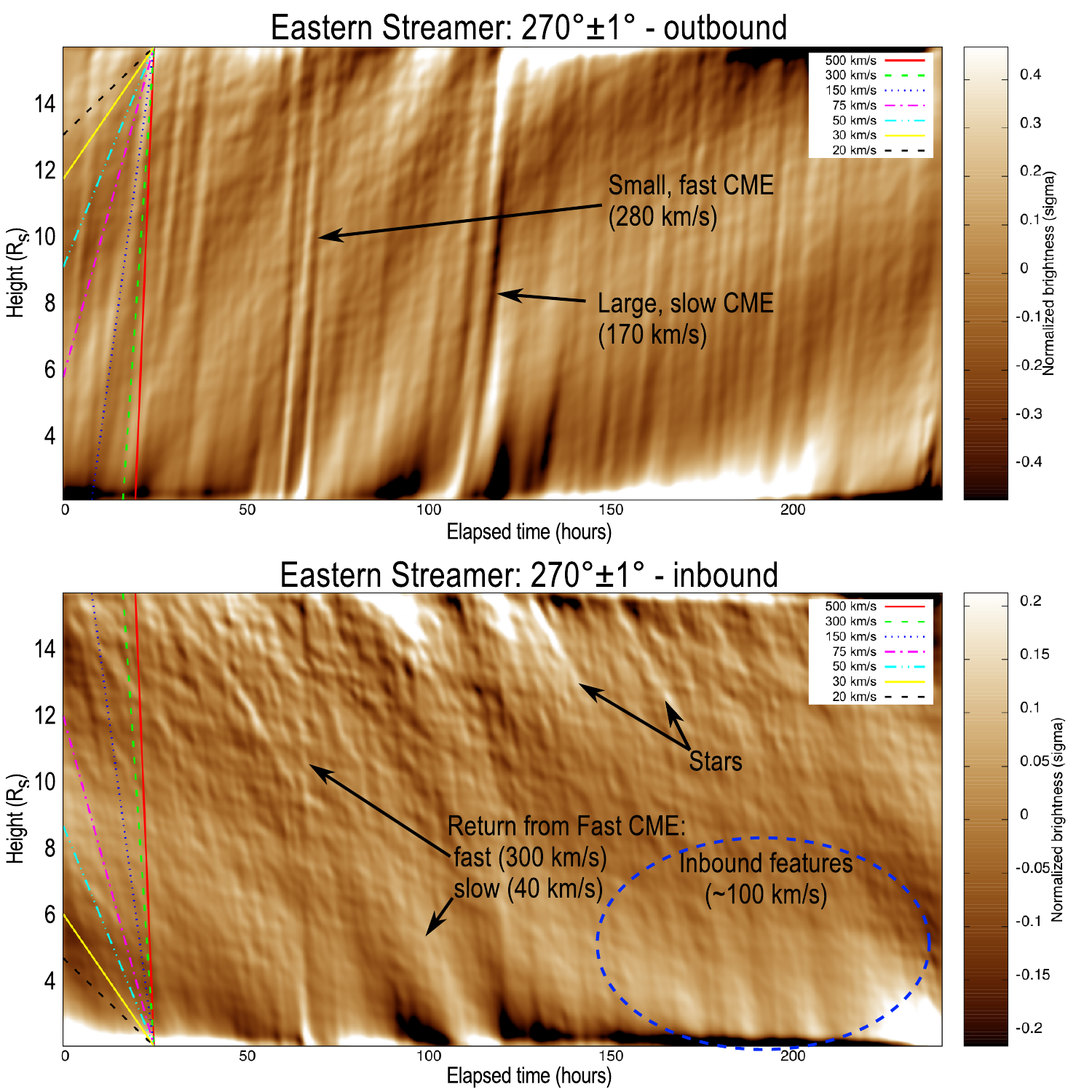}}
\protect\caption{\label{fig:streamer-diagram}Time/radius diagrams of filtered, separated
COR2 data at azimuth=270\textdegree{} show inbound and outbound features
in the streamer belt, including stars, CMEs, and an inbound CME return
signal. Speed is represented as slope in these images. Several fiducial
speeds are plotted as overlain lines.}
\end{figure*}

The fully filtered data form a pair of data cubes with independent
variables of azimuth, radius, and time. Figure \ref{fig:Fully-filtered-COR2}
is a constant-time slice of the two data cubes, at 2007 August 8 21:37
UT. It is more instructive to view a particular azimuth slice, plotting
filtered radiance against time and radius. Figure \ref{fig:streamer-diagram}
is such a slice, averaged over 2\textdegree{} of azimuth. The averaging
further beats down noise in the original data, and was chosen to match
the observed 2\textdegree{} size of coronal features such as polar
plumes (e.g. \citealt{FisherGuhathakurta1995}). Again, the full dataset
is available in the digital version of this article, as a movie that
runs over azimuth. Viewing the data in this way reveals azimuthal
structure in the corona. 

\begin{figure*}
\center{\includegraphics[width=5.0in]{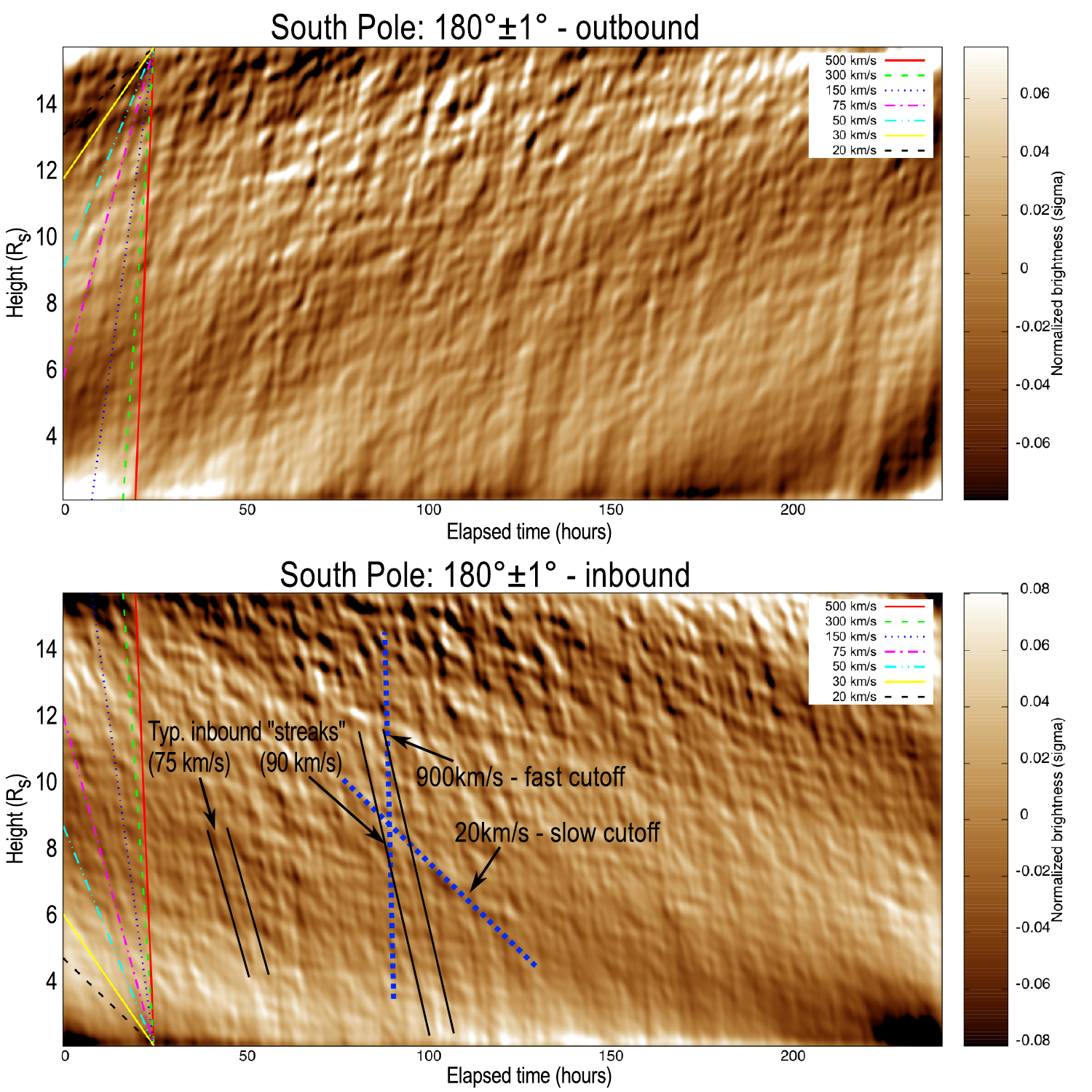}}
\protect\caption{\label{fig:polar-diagram}Time/radius diagrams at azimuth=180\textdegree{}
show inbound and outbound features over the polar coronal hole. The
inbound features are distinguishable from noise by their well-defined
characteristic speed, far from the cutoff speeds of the motion filter. }
\end{figure*}

\section{\label{sec:Results}Results}

Both Figure \ref{fig:streamer-diagram} and Figure \ref{fig:polar-diagram}
show evidence of inbound features in the solar corona. Each figure
has outbound features in the top panel and inbound features in the
bottom panel. Each panel's brightness scale is set to \textpm 2.5
times the calculated variance of the filtered radiance, and the mean
value has been subtracted from the image. Because of the radial normalization,
the motion-filtered radiance is in units of the measured pixel-value
variance at each radius from the Sun. For example, a value of +0.1
indicates a feature that is brighter than the mean radiance at its
radius from the Sun, by 10\% of the RMS variation of the original
pre-filtration data at that radius; and a pixel value of -0.01 indicates
a feature that is fainter than the mean by 1\% of that RMS variation.

Because of the complexity of the dynamics of the streamer belt, and
the relative simplicity and faintness of the coronal hole, we analyze
and report results from those two portions separately. In the streamer
belt there are sufficient visually distinguishable features to demonstrate
inbound wave motion from particualr excitation events; in the coronal
hole, it is both necessary and possible to perform speed-spectrum
analysis of the optical flow in the scene.

\subsection{Streamer Belt}

Figure \ref{fig:streamer-diagram} has several important outbound
features. Two CMEs erupted from the eastern streamer during this observation,
and are visible in the movies that accompany the online edition. The
first, at about 60 hours from the start of the data set, was a small
CME traveling at 280\textpm 20 km s$^{-1}$ between 8 and 14 $R_{S}$,
based on direct measurement of the feature's slope in the plotted
image. The second, at about 110 hours from the start of the data set,
was a larger CME traveling at a slower speed of 170\textpm 15 km s$^{-1}$
across that height range. Throughout the sequence, small outbound
features may be seen propagating at speeds from 150-400 km s$^{-1}$;
these appear to be the familiar ``blobs'' analyzed by \citet{Sheeley2009}. 

Inbound features in Figure \ref{fig:streamer-diagram} include several
residual star tracks, annotated in the figure; myriad diffuse inbound
features in the lower corona, four of which are circled between 150-250
hours; and returning inbound features from the first CME. Two clear
inbound features may be seen. First, a small, compact, bright inbound
feature may be a fast mode wave or retracting loop. The two are both
expected to propagate at the fast speed $V_{f}\sim V_{A}$, and the
observed feature speed is $\sim300$ km s$^{-1}$. The second is a
more diffuse, bright inbound feature moving at $\sim40$ km s$^{-1},$
which is consistent with a slow speed $V_{s}\sim C_{s}\sim100$ km
s$^{-1}$, slowed by outbound bulk motion of order 50 km s$^{-1}$.\emph{
}Both return signatures intersect the outbound CME
in the range 12-14 $R_{S}$. 

Several other bright, easily distinguished inbound features are present
and annotated in Figure \ref{fig:streamer-diagram}, propagating at
speeds between 40-100 km $s^{-1}$, and the compact, bright, slower-moving
star tracks in the outer portion of the image. Because these features
are easily recognized by eye, and are present with essentially constant
inbound speed at altitudes as high as 12-13 $R_{S}$, we immediately
conclude that the Alfvén surface is at least that high; this result
is refined in Section 5.

Superposed on the large-scale pattern is a lower-amplitude, more complex
background signal that is present at all azimuths. Because this background
signal is present both in the streamer belts and in the coronal holes,
where it is not mixed with the larger-scale evolution of the CMEs
and blobs, we analyze it primarily in the coronal holes.

\subsection{Coronal Hole}

Figure \ref{fig:polar-diagram}, being from the center of the southern
polar coronal hole, lacks the large-scale structures evident in the
streamer belt. It is thus simpler and easier to distinguish the
background signal, which has a complex character that at first glance
is difficult to distinguish visually from noise. There is a strong
characteristic speed to the background, as evidenced by the long,
narrow appearance of individual fluctuations. These have a
characteristic inbound speed of 40-90 km s$^{-1}$, varying across the
data set, which may be read directly from the typical slope of the
long, narrow fluctuations.  The lower speed bound of $\sim 40 km
s^{-1}$ is $\sim 2\times$ faster than the slow cutoff speed of the
processing, indicating it is not an artifact of the motion filter.
The corresponding characteristic outbound speed of the fluctuations in
the outbound panel is 200-400 km s$^{-1}$. These speeds are consistent
with an outbound subsonic wind with fast-mode Mach number in the range
0.5-0.8 assuming that the average between the typical inbound and
outbound velocities represents the bulk speed and the difference
represents twice the wave speed.

The structure of the inbound features, in particular, is important
for distinguishing them from noise. Typical long, narrow features
may be traced through 4-5 $R_{s}$ of inbound motion compared to their
instantaneous radial sizes of under 0.5 $R_{s}$. Typical features
span over 10 hours of elapsed time. This degree of elongation distinguishes
them from noise. Noise features may be expected to be incoherent on
timescales comparable to the instantaneous size of the feature, divided
by the difference between the two cutoff speeds of the motion filter
-- i.e. 1-2 hours for features similar to the annotated one. The characteristic
speeds of the motion filter are shown in Figure \ref{fig:polar-diagram}
to illustrate that the fluctuations' typical speeds are both well
defined and well between the filter cutoff speeds. In particular,
incoherent noise filtered through our motion filter would produce
``bowtie'' features with similar opening angle to the two filter
speeds. The observed fluctuations are more coherent.

\begin{figure}
\begin{centering}
\includegraphics{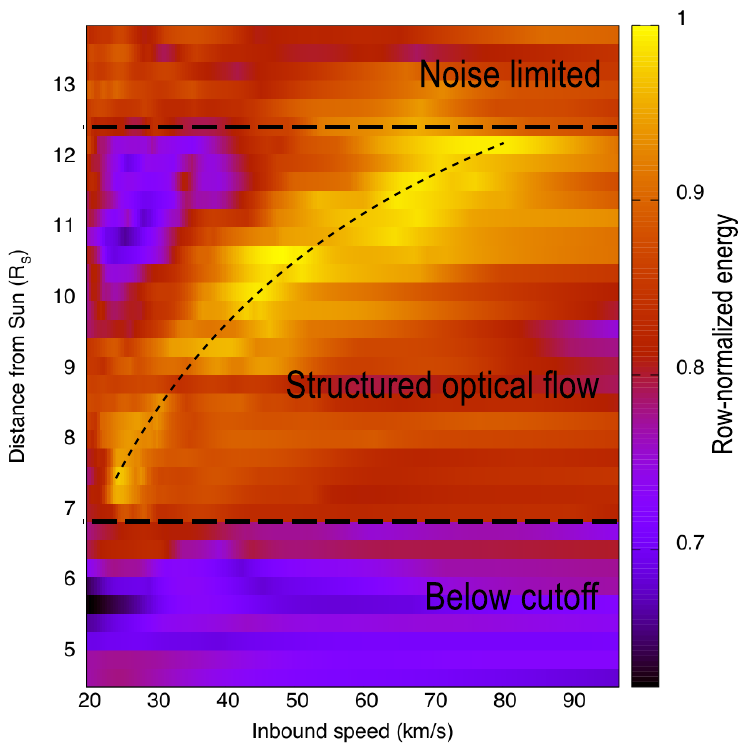}
\par\end{centering}

\protect\caption{\label{fig:Inbound-speed-spectra}Inbound speed spectrum vs. altitude
in the polar region shown in Figure \ref{fig:polar-diagram} reveals
structure in the seemingly random fluctuations of inbound coronal
radiance. The fluctuations have a well-defined characteristic speed
that varies with height; we take this as evidence that the fluctuations
are inbound waves, propagating with the upstream wave speed $V_{A}-V_{w}$;
and that the Alfvén surface is well above the 12 $R{}_{S}$ noise
limit of the present observation.}
\end{figure}

To better characterize the speed of the inbound features and to demonstrate
that they are not noise, we prepared a speed spectrum vs. altitude
over the south pole of the Sun. We prepared this spectrum by selecting
a $50$ h $\times$ 1.2 $R_{S}$ region of Figure \ref{fig:polar-diagram}
and convolving it with a diagonal line at a particular speed, then
calculating the RMS value of the convolved image. This RMS value formed
a single pixel of a (speed,radius) planar image, and we iterated over
both speed and central radius from the Sun of the extracted patch.
After generating all the RMS values, we normalized each row (i.e.
constant-radius locus) in the image to set its maximum value to unity.
We repeated the entire process for 5 randomly chosen 50 h intervals,
and collated a single image out of the median value of each pixel
over those 5 intervals. The result is the speed spectrum shown in
Figure \ref{fig:Inbound-speed-spectra}. The inbound features form
a well-defined, if slightly broadened, ridge relating speed and distance
from the Sun. The coherence of the ridge indicates the presence of
well-formed inbound movement at all radii out to 12 $R{}_{S}$.
Above 12 $R{}_{S}$ a clear ridge is not present, which likely indicates
that the noise floor dominates the measurement above that altitude.
Below 7 R$_{S}$, no clear ridge is visible -- but this is not surprising,
because the extrapolated speed of the ridge would be below the motion-filter
cutoff speed imposed during preprocessing.$ $

Because of the well-defined speed spectrum ridge, we identify the
polar inbound features as inbound waves. We would not expect clear
structure in the accidental speed profile of an ensemble of individual
packets of plasma that had been accelerated by different events; but
such structure is in fact expected for waves, whose speed is controlled
by the mostly-uniform medium that supports them. We identify the characteristic
speed as the upstream wave speed $V_{A}-V_{wind}$, and note that the
generally increasing trend in inbound speed with altitude indicates
that the wave speed was increasing faster with altitude than was the
wind speed. 

As a check on our intepretation of the speed data, we compare the
ridge speeds from Figure \ref{fig:Inbound-speed-spectra} to rough
estimates of the wind and Alfvén speed. Taking the average Alfvén-open
flux density to be $\sim$10 G at the photosphere (\citealt{DeForest1997}), and the linear expansion factor to be $\sim$2 (\citealt{DeForest2001b}),
the average magnetic field strength at 10 $R_{S}$ is 20 mG. Taking
the typical coronal hole electron density at 10 $R_{S}$ to be $n_{e}=$2$\times10^{4}$
cm$^{-3}$ (\citealt{DeForest2001a}) yields an Alfvén speed
$V{}_{A}\approx$300 km s$^{-1}$. Typical modeled solar wind speeds
at this altitude are in the range of $V_{w}\approx$ 200-300 km s$^{-1}$
(e.g. \citealt{Cranmer2013}), which is consistent with the interpretation
that the ridge is formed by waves moving at the upstream speed $V_{A}-V_{w}$. 

\section{\label{sec:Discussion}Discussion}

The primary physical difference between the solar corona and the solar
wind is the presence of inward-propagating wave signals. By separating
coronal features by inbound vs. outbound optical flow direction, we
have demonstrated, for the first time, remote measurement of these
inbound waves in the outer solar corona above 5 $R_{S}$. Based on
the observed properties of inbound features, we have determined a lower limit
for the Alfvén surface altitude of 15 $R_{S}$ in the streamer belt
and 12 $R_{S}$ in the coronal hole. The height of the lower bound
is set in the streamer belt by the field of view of the instrument,
and in the coronal hole by the noise properties of the particular
data set we used. 

There is evidence that the polar Alfvén surface is much higher than 12
R$_{S}.$ The propagation speed of the inbound waves appears to
\emph{increase} with height, which is the opposite direction of
variation from our \emph{a priori} expectations. Near the Alfvén
surface, the inbound wave speed must drop to zero with increasing
altitude, and the fact that the ridge in Figure
\ref{fig:Inbound-speed-spectra} is still increasing in speed as it
reaches the noise floor at $12R_{S}$ implies a further significant
height range in which the inbound speed decreases smoothly to zero. We
surmise that a smooth transition to zero would require at least a few
$R_{S}$, and therefore that the Alfvén surface must be at least
several $R_{S}$ above the top of the observed ridge in Figure
\ref{fig:Inbound-speed-spectra}. Deeper coronal exposures, and/or a
wider field of view, are necessary to extend the measurement further
from the Sun. It is neither necessary nor expected that the Alfvén
surface will prove to be spherical, smooth, or time-invariant.

As with all purely image-based measurements of motion, the data themselves
cannot directly distinguish between bulk motion and wave motion. In
our streamer belt analysis, this ambiguity is particularly keen in
the case of the downward signature of the first CME in Figure \ref{fig:streamer-diagram}.
Fortunately for the present study, both bulk retraction of a loop
and a downward propagating density wave are limited by the fast-mode
speed of the medium: we are able to use the retraction to place a
lower limit for the Alfvén surface location, regardless of whether
it be a tension-force-driven bulk motion or a pure inbound wave. The
inbound fluctuations observed in the coronal hole are wholly new and
identification is important to understand the phenomenon being observed.
The coherence of individual fluctuations in Figure \ref{fig:polar-diagram}
indicates that the features are not noise. The observed smoothly-varying
preferred speed versus radius strongly indicates wave motion, because
individual features would be expected to have a broader range of speeds
and no coherent ridge structure. The ridge pattern indicates that
the motions are governed by bulk properties of the corona rather than
by the accidental circumstances of formation of myriad small dense
objects, and the obvious bulk properties are the general outflow speed
and MHD wave speeds.

The Alfvén surface is important both as a boundary of the corona and
because of its important topological properties with regard to the
magnetic field. Magnetic field lines that exist entirely inside the
Alfvén surface are ``Alfv\'en closed'' and can move up or down through the
corona, or even in principle disappear entirely if their footprints
in the photosphere merge in the process of cancellation, which is
associated with submergence of magnetic flux under the photosphere
(e.g. \citealt{Schrijver1997,DeForest2007}). Field lines that penetrate
the Alfvén surface are ``Alfv\'en open'' in the sense that they cannot retract
into the Sun. A coronal loop, CME, or connected bolus of ejecta that
travels beyond the Alfvén surface must necessarily increase the heliospheric
magnetic flux, because the particular field lines now connect the
Sun to the heliosphere. Because it is impossible to retract these
Alfvén-open field lines, balancing the insertion of new magnetic flux
through the Alfvén surface into the heliosphere requires disconnection
of Alfvén-open field lines and subsequent ejection of U-shaped loops
outward through the Alfvén surface (\citealt{McComas1992,McComas1995,Schwadron2010}).
Reconnection above the Alfvén surface cannot affect the coronal flux
balance or morphology; reconnection below the Alfvén surface is necessary
to prevent the interplanetary magnetic field from growing without
limit.

Detecting and measuring the wave field in the outer corona is an important
step toward using the wave field as an independent measure
of solar wind acceleration and other coronal properties throughout
the outer corona. With better noise levels, it should be possible
to map the Alfvén surface directly. Even more importantly, it should
be possible, with improved noise levels and a custom observing campaign,
to measure the wind speed and outbound wave speed directly in the
coronal hole. Measuring all three, and incorporating photometric measurements
of the coronal density, will enable independent determination of $V_{A}$
and the magnetic field $B$ across height throughout the important
acceleration region of the solar wind.

\section{\label{sec:Conclusions}Conclusions}

We have, for the first time, detected inbound compressive waves in the
outer solar corona, and used them to set a strong lower limit on the
location of the Alfvén surface that marks the top of the solar corona
and beginning of the solar wind. We accomplished this measurement by
separating inbound and outbound density features through a Fourier
transform analysis of existing synoptic coronagraph data. We find that
the Alfvén surface was above 15 $R_{S}$ in the streamer belt and
significantly above 12 $R_{S}$ in the polar coronal holes.  These
limits imply that the upcoming Solar Probe Plus mission planned by
NASA to plunge repeatedly within 10 $R_{S}$ should routinely observe
the subsonic solar wind and corona in-situ.  These limits are set by
the field of view of the instrument, and the noise characteristics of
this particular measurement, respectively.  This initial detection and
measurement of the inbound wave field is an important first step
toward direct measurement of plasma properties throughout the entire
solar wind acceleration region.

\acknowledgements{This work was supported under grant from NASA's Heliophysics SHP-SR
program. D. Mc. was supported via the SWEPAM instrument on NASA's
ACE mission. The authors gratefully acknowledge the STEREO team for
making their data available to the public, and Marco Velli for illuminating
and helpful discussion. The analysis relied heavily on the freeware
Perl Data Language (http://pdl.perl.org).}

~

\pagebreak

\section{\label{sec:Appendix}Appendix: Motion filtering and speed spectrum analysis}

In this Appendix we describe some of our processing techniques, which
may be unfamiliar to the casual reader.  In Section 7.1,
we describe the basis of Fourier motion
filtering, how it works, and why we use it.  In Section 7.2,
we relate the RMS motion filter that we used to characterize
inbound speed vs. altitude over the coronal hole, to the motion filtering
described in Section 7.1.

Here we consider images as real or complex functions, mapping
$\mathbb{R}^2\rightarrow\mathbb{R}$ or
$\mathbb{R}^2\rightarrow\mathbb{C}$ as appropriate.  Actual digital
images are better represented on the integers, mapping 
$\mathbb{Z}^2\rightarrow\mathbb{C}$, but the arguments hold for both
cases.

\subsection{\label{sec:motion-filtering}Motion filtering with the Fourier transform}

The Fourier transform has many useful properties for image
transformation (\citealt{Bracewell}); here we use it for the property
of localizing moving features.  In particular, a 2-dimensional Fourier
transform in the $(r,t)$ plane localizes all features that are moving
with speed $v$ to a line of slope $-v$.  Thus, the three-step operation of
(i) Fourier transforming an $(r,t)$ image to its conjugate
$(k_{r},\omega)$ plane; (ii) filtering the resulting image to keep only
pixels with a certain range of the ratio $\omega k_r^{-1}$; and (iii) inverse
Fourier transforming back to the $(r,t)$ plane has the effect of
retaining only features moving within the corresponding range of
speeds $v = -\omega k_r^{-1}$.  The process removes all other features from the 
final $(r,t)$ image.

Here we demonstrate for the careful but unfamiliar reader that
features moving at a given velocity are indeed localized by the
Fourier transform, so that they can be retained or removed by masking
the Fourier plane.  Consider a time-distance image $I\left(r,t\right)$
that maps the value of some quantity as a function of position ($r$)
and time ($t$).  Take, as an \emph{ansatz}, that any
$I\left(r,t\right)$ can be decomposed by velocity:
\begin{equation}
I\left(r,t\right) = \int{ f_v(s_v) dv }\label{eq:decompose-by-velocity}
\end{equation}
where each $f_v$ is a separate function of a single variable $s_v$, with:
\begin{equation}
s_v \equiv\ r - v t \label{eq:def-of-s}.
\end{equation}
One obvious example of an $f_v(s_v)$ is an infinite plane wave
propagating at speed $v$, but there is no reason to consider only
plane waves.  For the following analysis literally any physically
relevant function $f(s)$ will suffice.  The definition of $s_v$
ensures that the corresponding $f_v(r,t)$ propagates the pattern at
the correct speed.

Clearly, if the \emph{ansatz} holds, then it is sufficient to
demonstrate that the 2-D Fourier transform localizes all signal energy from
just one single $f_v\left(s_v(r,t)\right)$ to a line $\omega = - v k_{r}$: 
since the Fourier transform is a linear operator, the
integral in Equation \ref{eq:decompose-by-velocity} migrates through
the transform operator.

The 2-D Fourier transform of $f_v(s_v)$ is just:
\begin{equation}
F_v(k_{r},\omega) \equiv \iint{e^{-ik_{r}r}e^{-i\omega t} f_v( r - v t ) dr dt}\label{eq:ft-of-fv}, 
\end{equation}
Switching variables to s in favor of r yields:
\begin{equation}
F_v(k_{r},\omega) = \iint{e^{-ik_{r}(s + vt)}e^{-i\omega t} f_v(s) ds dt}\label{eq:ft-in-s},
\end{equation}
which is easily separable because $f_{v}$ depends only on $s$ in this formulation.  Evaluating the $t$ integral:
\begin{equation}
\small{F_v(k_{r},\omega) = \left(2\pi\right)^{-1/2} \delta\left(vk_{r} + \omega\right) \int{ e^{-ik_{r}s}f_v(s) ds} }\label{eq:separated}
\end{equation}
where $\delta$ is the Dirac delta.  Clearly, $F_v(k_{r},\omega)$ is zero
everywhere except where $\omega = -k_{r}v$, i.e. the line of slope $-v$
in the $(k_{r},\omega)$ plane.  We have demonstrated that the moving
pattern $f_v(s_v(r,t))$ is localized to a particular line by the Fourier
transform, without any regard for the actual structure of the function 
$f_v(s)$.

Furthermore, because of the well-known invertibility of the Fourier
transform, and the linearity of both the Fourier transform and its
inverse, it is easy to see that the \emph{ansatz} must be
true. Clearly any function $F(k_{r},\omega)$ may be written as an
integral over $v$ of separately-defined $F_v(k_{r},\omega)$'s, since
the Dirac delta in Equation \ref{eq:separated} serves to isolate the
values of a particular $F_v(k_{r},\omega)$ from those of every other
$F_{v'}(k_{r},\omega)$, while $v$ spans the entire additional
dimension. Therefore every point $(k_{r},\omega)$ corresponds to a
specified (potentially nonzero) value of some $F_{v}(k_{r},\omega)$.
Since any function $f(r,t)$ may be represented as the inverse Fourier
transform of some $F_v(k_{r},\omega)$, the \emph{ansatz} holds. In
fact, the transform in Equation \ref{eq:decompose-by-velocity} is a
variant of the well-known \emph{radon transform}, of which speed
spectrum analysis is but one application.  The radon transform is
explored in some detail in Chapter 8.8 of \citet{Bracewell}.

In short, every image $I(r,t)$ may be represented as the integral over
$v$ of a collection of speed-filtered images, each of which contains only features
moving at particular speed $v$.  These components are localized in the
Fourier plane, and Fourier transformation can be used to isolate them.
Speed filters of these types have many applications.  In heliophysics,
applications include separation of solar photospheric p-modes with phase
speeds well above the local sound speed, from surface features that
move under the local sound speed (R. Shine, priv. comm. 1999;
\citealt{Lamb2010}); and isolation of solar wind features from quasi-stationary
artifacts \citep{DeForest2011}.  Interested readers are directed to 
Chapter 8 of \citet{Bracewell} for a range of fascinating insights.

In addition to the obvious benefit of isolating inbound and outbound
features from the data, motion filtering with a narrow range of speeds
also reduces photon noise in the resulting processed data.
Uncorrelated noise, such as photon shot noise, is distributed evenly
throughout Fourier space.  Zeroing out pixels in the $(k_r,\omega)$
plane to reduce the total nonzero pixel count by a factor of $\alpha$
thus reduces the total photon noise in the final $(r,t)$ image by a
factor of $\alpha^{1/2}$.

\subsection{\label{sec:rms-analysis}Speed spectrum analysis by convolution}

To identify a pattern of inbound energy versus radius over the South
pole of the Sun, we use convolution of an image patch in the $(r,t)$
plane, with a diagonal line.  We convolved the image patch with a
diagonal line of specified slope $v$, and took the RMS value of the
resulting image patch as an indication of the total number of features
moving at approximately speed $v$ in the patch. 

By varying both $v$ and the central radius $r_{cen}$ of the patch, we
were able to arrive at a map showing the relative distribution of
total integrated feature strength versus $v$ and $r$. This is the
incoherent variant of the well-known radon transform
(\citealt{Bracewell}, Chapter 8.8).

This process is a quick and easy way to identify patterns in the speed
spectrum of inbound waves.  In Section 7.1
we
demonstrated that a speed spectrum exists -- i.e. that any given
$(r,t)$ image can be represented as a collection of images, each of
which contains only features moving at a particular velocity $v$. In our
case we did not want to represent the individual features, only to 
estimate the total inbound image energy moving at a given speed, as a function
of that speed.

Here we demonstrate that convolution with a diagonal line is
equivalent to applying a speed filter in Fourier space as described in
Section 7.1.
Recall the famous \emph{Convolution Theorem} (e.g. Section 4.17 of \citealt{Bracewell}) that relates
convolution in real space to multiplication in Fourier space, and vice
versa:
\begin{equation}
f(r,t) * q(r,t) = \mathcal{F}^{-1}\left( F(k_{r},\omega) \otimes Q(k_{r},\omega) \right)\label{eq:conv-theorem}
\end{equation}
where $\mathcal{F}$ represents the Fourier transform, $F\equiv\mathcal{F}(f)$, $Q\equiv\mathcal{F}(q)$, $*$ is the
convolution operator, and $\otimes$ is elementwise multiplication.

Consider the ideal straight-line image: 
\begin{equation}
q_v(r,t) = \delta(r - vt)\label{eq:straight}.
\end{equation}
Its Fourier transform $Q_{v}\equiv\mathcal{F}(q_v)$ is easily calculated:
\begin{equation}
Q_{v}(k_r,\omega) \equiv \iint{ e^{-i\omega t}e^{-ik_{r}r} \delta(r - vt) dr dt}\label{eq:Qq},
\end{equation}
which may be performed by inspection since one of the two integrals is done by the $\delta$ and the other itself yields a $\delta$:
\begin{equation}
Q_{v}(k_r,\omega) = \left(2\pi\right)^{-1} \delta(k_{r}v + \omega)\label{eq:Qq2},
\end{equation}
which is a single-speed filter in the Fourier plane. Equation
\ref{eq:Qq2} should not be a surprise, since $q_v(r,t)$ matches the
form of \ref{eq:def-of-s} in Section 7.1.

Hence, convolution with a straight line is equivalent (up to a
multiplicative constant) to Fourier filtering with a perpendicular
straight line.  By neglecting to propagate constant values in
convolving our image patches with various straight lines, we lost any
photometric quality to the remaining RMS value of the image patch
after filtration -- but as we were interested in detecting a pattern in
the surviving optical flow, rather than in quantifying the total image
energy in that flow, simple convolution was sufficient.

\end{document}